\documentclass{svproc}
\usepackage{xcolor}
\usepackage{amssymb}
\usepackage{amsmath}
\usepackage{multirow}
\usepackage{arydshln}
\usepackage[switch]{lineno} 
\usepackage[misc]{ifsym}

\newcommand{\comment}[1]{}

\usepackage{collcell}
\usepackage{booktabs}
\usepackage{siunitx}
\usepackage{etoolbox}
\usepackage{ifthen}
\usepackage[caption=false]{subfig}
\usepackage{rotating}
\usepackage{changepage}
\usepackage{cancel}
\usepackage{caption}
\usepackage{wasysym}
\usepackage{arydshln}
\usepackage[acronym,toc]{glossaries}
\usepackage[normalem]{ulem}
\usepackage{tabularx}

\usepackage{shorttoc}
\usepackage[subfigure]{tocloft}
\usepackage{hyperref}

\robustify\bfseries

\usepackage{url}

\begin{document}
\mainmatter              % start of a contribution
\title{Properties of Reddit News Topical Interactions}
\titlerunning{Reddit News Topical Interactions}  % abbreviated title (for running head)
%                                     also used for the TOC unless
%                                     \toctitle is used
%
\author{Ga\"el Poux-M\'edard\inst{1} \and Julien Velcin\inst{1} \and Sabine Loudcher\inst{1}}
\authorrunning{Ga\"el Poux-M\'edard et al.} % abbreviated author list (for running head)
%
%%%% list of authors for the TOC (use if author list has to be modified)
\tocauthor{Ga\"el Poux-M\'edard, Julien Velcin, and Sabine Loudcher}
\institute{Université de Lyon, Lyon 2, ERIC UR 3083, Bron 69676, France\\
\email{gael.poux-medard@univ-lyon2.fr},\\ WWW home page:
\texttt{https://gaelpouxmedard.github.io}}

\maketitle              % typeset the title of the contribution

\begin{abstract}
Most models of information diffusion online rely on the assumption that pieces of information spread independently from each other. However, several works pointed out the necessity of investigating the role of interactions in real-world processes, and highlighted possible difficulties in doing so: interactions are sparse and brief. As an answer, recent advances developed models to account for interactions in underlying publication dynamics. In this article, we propose to extend and apply one such model to determine whether interactions between news headlines on Reddit play a significant role in their underlying publication mechanisms. After conducting an in-depth case study on 100,000 news headline from 2019, we retrieve state-of-the-art conclusions about interactions and conclude that they play a minor role in this dataset.
\end{abstract}

\section{Introduction}
As the volume of data available on the Internet keeps on growing exponentially, so does the need for efficient data-processing tools and methods. In particular, the user-generated content produced on online social platforms provides a detailed snapshot of the world population's thoughts and interests. This kind of data can be used for many different applications (e.g. in marketing, opinion mining, fake news control, summary generation, etc.). However, we need a fine understanding of the underlying data-generation mechanisms at stake to refine the results of these possible applications.

\begin{figure}[t]
    \centering
    \includegraphics[width=\textwidth]{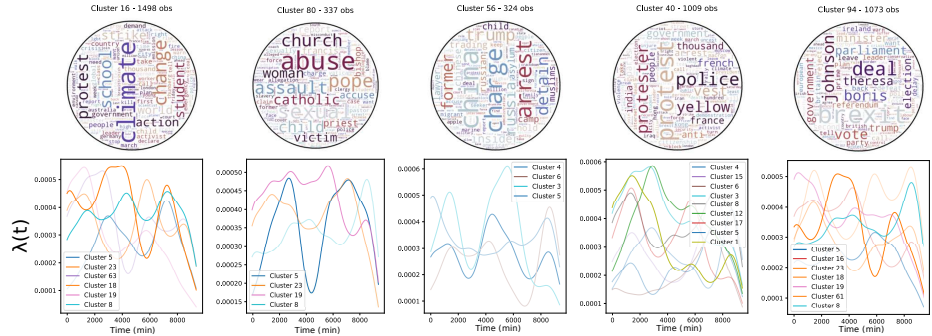}
    \caption{\textbf{An output of the proposed approach} -- (Top) A set of inferred topics along with the vocabulary of their documents. (Bottom) Instantaneous probability for one observation of a topic to trigger other observations in other topics.}
    \label{fig-illustrMPDHPClus}
\end{figure}
Early works on information spread consider that a user publishing a piece of information (or meme) did so due to an earlier exposure to this same meme \cite{Kempe2003IndependentCascade}. Users are represented as a network on which edges pieces of information flow independently from each other. This model has seen several refinements, that consider additional information about nodes and edges to model the way information spreads. Nodes participate the spread when exposed a certain number of times \cite{Myers2012CoC,Niemczura2015LinThresMod}, have a different influence on their neighbours depending on their position in the network \cite{Larremore2012StatisticalPropertiesAvalanches}, can publish exogenous memes (e.g. that were not flowing on the network beforehand) \cite{Myers2012ExternalInfluence,He2015HawkesTopic}, etc.. Edges between nodes represent the likeliness of a meme flowing from one node to the other; later works consider edges whose intensity depends on the meme considered \cite{Du2013TopicCascade,Barbieri2017SurvivalFactorization}, on the temporal distance to the last exposure \cite{GomezRodriguez2011NetRate,GomezRodriguez2013InfoPath}, etc.

However, a core assumption of most of these models and extensions is that memes spread independently from each other. It has been highlighted in some occasions that modeling the interaction between pieces of information might bring interesting insights in the way data gets generated online \cite{Beutel2012InteractingViruses,Myers2012CoC,Rodriguez2013StructureAD}. For instance, we expect memes about two politically opposed candidates to have some interaction (the publication of one would inhibit the publication chances of the other one); memes raising climate change awareness might be more likely to spread when coexisting with memes about ecologic disasters, etc.

In this work, we conduct an in-depth study of such interaction mechanisms on a large-scale Reddit dataset --gathered from 9 news subreddits over 2019. We propose to use a model that answers the challenges of interaction modeling raised in recent works on interactions --interactions between memes are likely sparse and brief. This model, the Multivariate Powered Dirichlet-Hawkes Process (MPDHP), groups textual memes into topics. A possible output is represented in Fig.~\ref{fig-illustrMPDHPClus}. Topics interact in pairs to yield a probability for a new meme about a given topic to get published, in a continuous-time setting. The output of MPDHP is a time-dependent topical interaction network. Once this network has been inferred, we can analyse the properties of topical interactions. We conduct experiments using several iterations of MPDHP (accounting for different timescales and topical modeling), so that we get an exhaustive panorama of interactions at stake. Overall, our conclusions hint that interactions play a minor role on news subreddits; most information indeed spreads independently in this context. From a broader perspective, we introduce a methodology to assess the role of temporal topical interactions textual datasets.

\section{Background}
\subsubsection{Modeling interactions}
In this section, we briefly review previous works on interaction modeling in spreading processes. To the best of our knowledge, the first work tackling the problem proposes a SIR-based model that comprises an interaction term $\beta$ between two co-existing viruses --in their case, the adoption of either Firefox or Chrome web browser \cite{Beutel2012InteractingViruses}. Interaction is modeled as a hyper-parameter, that has to be tuned manually to retrieve global shares of each virus over time. Building on this work, \cite{Myers2012CoC} later proposed to model interaction between memes at the agent level: given a user is exposed to tweet A at time $t_A$, what is the probability to retweet tweet B at a time $t_B$ later. This model attempts to learn the interaction parameter $\beta$ from the data instead of fine-tuning it. Their conclusion is that interactions can have a large overall effect of the data-generation processes (here retweeting mechanisms), but that most interactions have little influence. Later works tackling the problem from a similar perspective on several real-world datasets find similar results: significant interactions between clusters of memes are sparse \cite{Poux2021IMMSBM,Poux2022InterInfoSpread}. This highlight the need to cluster memes so that it becomes possible to retrieve meaningful interaction terms.

Some other works tackled the problem of pair-interaction modeling from a temporal perspective. Simple considerations show that interactions cannot remain constant over time; a user that read a meme 5min earlier and another meme 5 days earlier is likely to be much more influenced the first one than the second one. Interactions are likely to be brief. This temporal dependence is explicitly modeled in \cite{Poux2021InterRate}, where the interaction strength is shown to typically decay exponentially, which correlated the findings of \cite{Cao2019AdsDataset}, and of \cite{Myers2012CoC} to a certain extent. It highlights the need to consider time to relevantly model interactions.
Thus the need for a model able to handle large piles of data, performing topic inference, allowing interactions between these topics, and accounting for time. 

\subsubsection{Dirichlet-Hawkes processes}
The Dirichlet-Hawkes process \cite{Du2015DHP} seem to qualify for these requirements. In particular, it has been extended as the Powered Dirichlet-Hawkes process \cite{Poux2021PDHP} to allows for extended modeling flexibility --to which extend should we favor the textual information over the temporal information. However, both these models do not account for pair-interaction between the inferred topics. Instead, topics can only trigger new observations from themselves. In this work, we consider a Multivariate extension of the Powered Dirichlet-Hawkes process (MPDHP). This extension boils down to substituting the Hawkes process described in \cite{Du2015DHP,Poux2021PDHP} by a multivariate Hawkes process \cite{Liniger2009MultivariateHawkesProcess}, so that topics can influence the probability of new observations to belong to either other topics.

The principle of the Dirichlet-Hawkes approaches relies on Bayesian inference. An online language model accounts for the textual content of documents (Fig.~\ref{fig-illustrMPDHPClus}-top). This model, typically a Dirichlet-Multinomial bag of words as we will use later, expresses the likelihood for a new document to belong to any existing topic. This language model is coupled to a Dirichlet-Hawkes prior, that can be expressed as a sequential process. The prior assigns to a new observation a prior probability to belong to either topic based on its temporal interaction with earlier publications (Fig.~\ref{fig-illustrMPDHPClus}-middle). In the remaining of this section, we present the modified formulation of the base models \cite{Du2015DHP,Poux2021PDHP} so that it accounts for topical pair-interactions.

We first rewrite and detail the expression of the Dirichlet-Hawkes process as introduced in \cite{Du2015DHP,Poux2021PDHP}:
\begin{equation}
\label{eq-MPDHP-prior}
\begin{split}
    P&(C_i=c \vert n_i, N_{<i,c}, t_i, \lambda(t_i), \mathcal{H}, \theta_0, \lambda_0, r) \\
    &\propto \underbrace{P(n_i \vert C_i=c, N_{<i,c}, \theta_0)}_{\substack{\text{Textual likelihood} \\ \text{(Dirichlet-Multinomial)})}} \underbrace{P(C_i = c\vert t_i, \lambda(t_i), \mathcal{H}, \lambda_0, r)}_{\substack{\text{Temporal prior} \\ \text{(MPDHP)}}} \\
    &= \frac{\Gamma(N_c+\theta_0)}{\Gamma(N_c+n_i+\theta_0)} \prod_v \frac{\Gamma(N_{c,v} + n_{i,v} + \theta_{0,v})}{\Gamma(N_{c,v}+\theta_0)} 
    \begin{cases}
    \frac{\lambda_c^r(t_i)}{\lambda_0 + \sum_{c'} \lambda_{c'}^r(t_i)} \text{ if c$\leq$K}\\
    \frac{\lambda_0}{\lambda_0 + \sum_{c'} \lambda_{c'}^r(t_i)} \text{ if c=K+1}
    \end{cases}
\end{split}
\end{equation}
where $C_i$ represents the cluster chosen by the $i^{\text{th}}$ document, $c$ is the random variable accounting for this allocation, $n_i$ is a vector that whose $v^{\text{th}}$ entry represents the count of word $v$ in document $i$, $N_{<i, c}$ is the vector whose entry $v$ represents the total count of word $v$ in all documents up to $i$ that belong to cluster $c$, $t_i$ the arrival time of document $i$, $\lambda(t_i)$ the vector of intensity functions at time $t_i$ whose $c^{\text{th}}$ entry corresponds to cluster $c$, and $\mathcal{H}$ the history of all previous documents that appeared before time $t_i$. The three last symbols are hyperparameters: $\theta_0$ is the concentration parameter of the Dirichlet-Multinomial language model \cite{Du2015DHP,Yin2018ShortTextDHP}, $\lambda_0$ is the concentration parameter of the MPDHP temporal prior \cite{Du2015DHP}, and $r$ controls the extent to which we rely of either textual of temporal information \cite{Poux2021PDHP}. 
Data is processed sequentially using a Sequential Monte Carlo (SMC) algorithm similar to \cite{Du2015DHP,Poux2021PDHP,Valera2017HDHP}. For each new observation, we get from Eq.~\ref{eq-MPDHP-prior} the posterior probability that it belongs to either of $K$ existing clusters, or to start a cluster of its own ($K+1$). The SMC algorithm accounts for several allocation hypotheses at once, and discard the most unlikely ones every other iteration. 

Now, the extension of the Dirichlet-Hawkes process to the multivariate case boils down to giving a new definition for the vector $\lambda(t)$. We express it as:
\begin{equation}
    \label{eq-lambda}
    \lambda_c (t) = \sum_{t_i^{c'} < t} \vec{\alpha}^T_{c,c'} \cdot \vec{\kappa}(t-t_i^{c'}) \,\,\,\,\,\,\,\text{where}\,\,\,\,\,\,\, \kappa_l(\Delta t) = \frac{1}{\sqrt{2 \pi \sigma_l^2}}e^{-\frac{(\Delta t-\mu_l)^2}{2 \sigma_l^2}}
\end{equation}
where $\vec{\alpha}_{c,c'}$ is a vector of $L$ parameters to infer, and $\vec{\kappa}(t-t_i^{c'})$ is a vector of $L$ given temporal kernel functions depending only on the time difference between two events. We consider $\vec{\kappa}(\Delta t)$ to be a Gaussian RBF kernel with fixed mean $\vec{\mu}$ and deviation $\vec{\sigma}$, which allows us to model a range of different intensity functions.
Each parameter $\alpha_{c,c',l}$ accounts for the influence of $c'$ on $c$ according to the $l^{th}$ entry of the temporal kernel. The dot product of $\alpha$ with $\kappa$ yields the intensity function vector $\lambda$, which represents the topical interactions' temporal adjacency matrix (represented in an alternative way in Fig.~\ref{fig-illustrMPDHPClus}-middle).

\section{Experimental setup}
\begin{figure}[t]
    \centering
    \includegraphics[width=\textwidth]{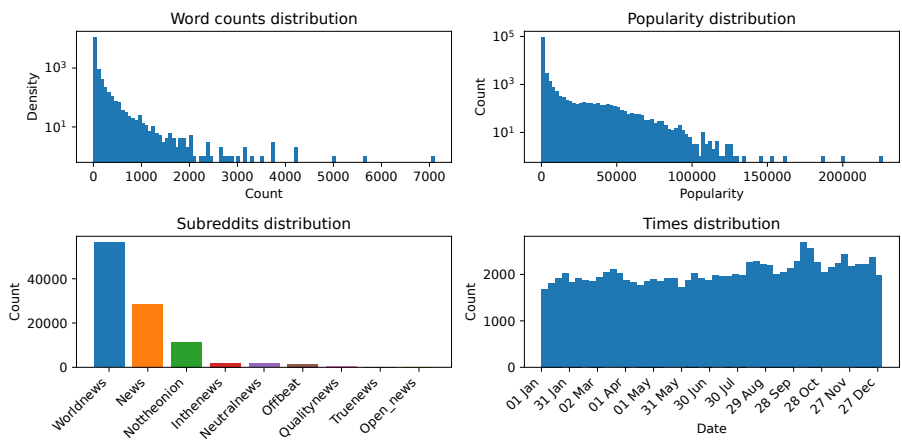}
    \caption[MPDHP x News - Characteristics of the News dataset]{\textbf{Characteristics of the News dataset} --- For $\sim$100,000 headlines and $\sim$13,000 different words: (Top-Left) Distribution of the words count. (Top-Right) Distribution of headlines popularity. (Bottom-Left) Distribution of headlines over subreddits. (Bottom-Right) Distribution of headlines over time.}
    \label{fig:statsDSRedditafter}
\end{figure}
\subsubsection{Dataset}
The dataset used in this study has been gathered from the Pushshift Reddit repository \cite{Baumgartner2020PushshiftRedditDataset}, which contains archives of the entirety of Reddit posts and comments up to June 2021. For each Reddit post, we can retrieve the subreddit it came from, the title of the publication, its publication date and its score (number of upvotes minus number of downvotes).

We choose to only consider popular English news subreddits. Namely, we select only posts from 2019 published on the following subreddits: inthenews, neutralnews, news, nottheonion, offbeat, open news, qualitynews, truenews, worldnews. This leaves us with a corpus of 867,328 headlines, which makes a total of 1,111,955 words drawn from a vocabulary of size 36,284. 

Finally, we discard uninformative words and documents from the dataset. Explicitly, we remove the stopwords, punctuation signs, web addresses, words whose length is lesser than 4 characters, and words that appear less than 3 times in the whole dataset. 
Then, we remove publications that carry lesser textual or temporal information. Firstly, we choose not to consider the publications that have a popularity lesser than 20 -- meaning that they received less than 20 positive votes more than negative votes. We make this choice so that we consider publications that are visible enough to have any influence on the data generation process. Secondly, we remove publications that comprise less than 3 words. The semantic information so-carried is expected to be poor and is not considered in our analysis.

After curating the dataset in the way described above, we are left with 102,045 news headlines (one-eighth of the original data), which makes a total of 875,334 tokens (named entities, verbs, numbers, etc.) drawn from a vocabulary of size 13,241 (one-third of the original vocabulary). The characteristics of this dataset are shown in Fig.~\ref{fig:statsDSRedditafter}. 

\subsubsection{Temporal kernel}
We run our experiments using three different RBF kernels, which account for publication dynamics at three different timescales: minute, hour, and day. 
The ``\textbf{Minute}'' RBF kernel is made of Gaussian functions centered at the following times: $\left[ 0, 10, 20, 30, 40, 05, 60, 70, 80 \right]$ minutes; each entry shares a same standard deviation $\sigma$ of 5 minutes, and $\lambda_0=0.01$.
The ``\textbf{Hour}'' RBF kernel has Gaussians centered around $\left[ 0, 2, 4, 6, 8 \right]$ hours, with a standard deviation $\sigma$ of 1 hour, and $\lambda_0=0.001$.
The ``\textbf{Day}'' RBF kernel is centered around $\left[ 0, 1, 2, 3, 4, 5, 6 \right]$ days, with a standard deviation $\sigma$ of 0.5 days, and $\lambda_0=0.0001$.
For each of these kernels, we set the concentration parameter $\lambda_0$ so that it reaches roughly the value of one Gaussian function evaluated at $2\sigma$. It means that an event which is $2\sigma$ away from the center of the Gaussian kernel of a single observation has 50\% chances of getting associated with this Gaussian kernel entry, and 50\% chances of opening a new cluster.

\subsubsection{Hyper-parameters}
We consider two values for the concentration parameter of the language model: $\theta_0=0.001$ and $\theta_0=0.01$.
The choice of this range is standard in the literature \cite{Du2015DHP} and supported by our own observations. A larger value of $\theta_0$ makes the inferred clusters cover a broader range of document types, whereas a small value makes the inferred clusters more specific to a topic.

The value of $r$ is chosen to be either $0$ (no use of the temporal information), $0.5$, $1$ and $1.5$. The larger $r$, the more the inference relies of the temporal dynamics instead of the textual content of the documents.

The SMC algorithm described in \cite{Du2015DHP,Poux2021PDHP} is run using 8 particles and 100,000 samples used to estimate the matrix of parameters $\alpha$.

\section{Results}
\subsection{Overview of the experiments}

\begin{table}[h]
    \centering    
    \caption{\textbf{Results for each experiment} --- We characterize each run in terms of clusters (number, size, textual entropy, subreddits entropy). The standard deviation on the last digits is given between parenthesis.}
    \label{tab-diffruns}
    \begin{tabularx}{\textwidth}{|>{\hsize=.25\hsize}X|>{\hsize=.25\hsize}X|>{\hsize=.5\hsize}X|X|X|X|X|}
        \cline{1-7}
        $\vec{\kappa} (t)$ & $\theta_0$ & $r$ & $K$ & $<N>$ & $S_{text}^{(20)}$ & $S_{sub}^{(20)}$ \\
        \cline{1-7}
        {\multirow{8}{*}{\rotatebox[origin=c]{90}{\textbf{Minute}}}} & {\multirow{4}{*}{\rotatebox[origin=c]{90}{\textbf{0.01}}}} & 0.0 & 16150 & 6 & 0.744(68) & 0.400(36) \\
        & & 0.5 & 8498 & 12 & 0.796(55) & 0.441(24) \\
        & & 1.0 & 5069 & 20 & 0.790(49) & 0.476(53) \\
        & & 1.5 & 2730 & 37 & 0.808(43) & 0.490(52) \\
        \cdashline{2-7}
        & {\multirow{4}{*}{\rotatebox[origin=c]{90}{\textbf{0.001}}}} & 0.0 & 46304 & 2 & 0.475(48) & 0.239(56) \\
        & & 0.5 & 37277 & 3 & 0.485(40) & 0.256(60) \\
        & & 1.0 & 26858 & 4 & 0.501(37) & 0.266(72) \\
        & & 1.5 & 19275 & 5 & 0.506(39) & 0.280(58) \\
        \cline{1-7}
        
        % ========================
        
        {\multirow{8}{*}{\rotatebox[origin=c]{90}{\textbf{Hour}}}} & {\multirow{4}{*}{\rotatebox[origin=c]{90}{\textbf{0.01}}}} & 0.0 & 3792 & 27 & 0.798(52) & 0.474(106) \\
        & & 0.5 & 1735 & 59 & 0.791(45) & 0.469(77) \\
        & & 1.0 & 825 & 124 & 0.803(47) & 0.484(75) \\
        & & 1.5 & 426 & 240 & 0.795(37) & 0.489(69) \\
        \cdashline{2-7}
        & {\multirow{4}{*}{\rotatebox[origin=c]{90}{\textbf{0.001}}}} & 0.0 & 18012 & 6 & 0.760(86) & 0.397(63) \\
        & & 0.5 & 11923 & 9 & 0.784(72) & 0.425(38) \\
        & & 1.0 & 4837 & 21 & 0.821(50) & 0.497(30) \\
        & & 1.5 & 2368 & 43 & 0.814(42) & 0.481(91) \\
        \cline{1-7}
        
        % ========================
        
        {\multirow{8}{*}{\rotatebox[origin=c]{90}{\textbf{Day}}}} & {\multirow{4}{*}{\rotatebox[origin=c]{90}{\textbf{0.01}}}} & 0.0 & 609 & 168 & 0.713(34) & 0.413(103) \\
        & & 0.5 & 326 & 313 & 0.728(33) & 0.429(98) \\
        & & 1.0 & 172 & 593 & 0.743(31) & 0.461(94) \\
        & & 1.5 & 96 & 1063 & 0.755(36) & 0.464(86) \\
        \cdashline{2-7}
        & {\multirow{4}{*}{\rotatebox[origin=c]{90}{\textbf{0.001}}}} & 0.0 & 4349 & 23 & 0.705(49) & 0.396(103) \\
        & & 0.5 & 2654 & 38 & 0.721(59) & 0.404(104) \\
        & & 1.0 & 1399 & 73 & 0.734(57) & 0.431(106) \\
        & & 1.5 & 764 & 134 & 0.741(54) & 0.442(98) \\
        \cline{1-7}
         
    \end{tabularx}
\end{table}

In Table~\ref{tab-diffruns}, we represent the main characteristics of each run in terms of number of inferred clusters $K$, the average cluster population $<N>$ (where $< \cdot >$ denotes the average), the average normalized entropy of the vocabulary of the top 20 clusters $S_{text}^{(20)}$, the average normalized entropy of the subreddits partition of the most populated 20 clusters $S_{sub}^{(20)}$.
The normalized entropy is bounded between 0 and 1. It is defined so that a low entropy $S_{text}^{(20)}$ (resp. $S_{sub}^{(20)}$) means that the top 20 clusters contain documents that are concentrated around a reduced set of words (resp. of subreddits); conversely, a large entropy means that these top 20 clusters do not account for documents concentrated around a specific vocabulary (resp. set of subreddits).
We can make several observations from Table~\ref{tab-diffruns}:
\begin{itemize}
    \item The number of inferred clusters decreases with $r$, and their average population increases.
    \item The number of clusters grows large for the ``Minute'' kernel. This is because the short time range considered does not allow for clusters to last in time. A cluster that does not replicate within 1h30 is forgotten.
    \item We recover the fact that textual clusters have a lower entropy for small values of $r$ \cite{Poux2021PDHP}; this is because their creation is based more on textual coherence than on temporal coherence.
    \item The subreddit entropy seems to increase as $r$ grows. A possible interpretation is that favouring the temporal information for cluster creation results in larger clusters (see $<N>$). They would be too large to account for subreddit-specific dynamics. However, the entropy remains lower than the entropy of the distribution Fig.~\ref{fig:statsDSRedditafter}-top-left, equal to 0.51.
\end{itemize}

%We recall that the normalized entropy is defined as:
%\begin{equation}
%    S(\vec{x}) = -\frac{1}{\ln \vert \vec{x} \vert}\sum_{i}^{\vert \vec{x} \vert} x_i \ln x_i
%\end{equation}
%where $\vec{x}$ is a vector that sums to $1$ and $\vert \vec{x} \vert$ its cardinal (length). Each entry $x_i$ represents the probability of $i$. When considering counts, $\vec{x}$ can be set equal to the frequency of each observation. The entropy is normalized between 0 (minimal spread, $\vec{x}_i = \delta_{ij} \, \forall i$) and 1 (maximal spread, $\vec{x}_i = \frac{1}{\vert \vec{x} \vert} \, \forall i$). In our case, a low entropy $S_{text}^{(20)}$ (resp. $S_{sub}^{(20)}$) means that clusters contain documents that are concentrated around a reduced set of words (resp. of subreddits); conversely, a large entropy means that clusters do not account for documents concentrated around a specific vocabulary (resp. set of subreddits).

\subsection{Quantifying interactions}
\subsubsection{Effective interaction}
We introduce the parameters we are going to use in follow-up analyses. The output of MPDHP consists of a list of clusters comprising timestamped bags of words --news headlines. Between each pair of clusters, MPDHP inferred a temporal influence function $\lambda (t)$, that represents the probability for one cluster to trigger publications from another. Therefore, our model yields an adjacency matrix $A \in \mathbb{R}^{K \times K \times L}$, where $K$ is the number of clusters and $L$ the size of the RBF kernel $\vec{\kappa} (t)$. One entry $a_{i,j,l}$ represents the strength of the influence of $j$ in $i$ due to the $l^{th}$ entry of $\vec{\kappa}(t)$.

However, we must consider the effective number of interactions, that is to which extent the intensity function has effectively had a role in the inference. To do so, we simply consider a weight matrix $W \in \mathbb{R}^{K \times K \times L}$, whose entries $w_{i,j,l}$ are the average of the intensity of $i$ above $\lambda_0$ due to $j$ from the kernel entry $l$ for all observations. Explicitly:
\begin{equation}
    \label{eq-effInter}
    w_{i,j,l} = \frac{1}{\vert \mathcal{H}_i \vert}\sum_{t_i \in \mathcal{H}_i} \sum_{t_j<t_i} max(a_{i,j,l} \kappa_l (t_i-t_j)-\lambda_0, 0)
\end{equation}
The notations are the same as in Eq.~\ref{eq-MPDHP-prior} and Eq.~\ref{eq-lambda}. Note that we retract $\lambda_0$ from the intensity term, because it is considered as a background probability for a publication to happen. Therefore, $W$ can also be interpreted as the instantaneous increase in probability due to interactions.

\subsubsection{Interactions strength}
In Table~\ref{tab-interStrength}, we investigate the effective impact of interactions in the dataset. We consider the following metrics:
\begin{itemize}
    \item $<A>$: the average value of the whole adjacency matrix; to which extent topics interact with each other according to MPDHP.
    \item $<W>$: the average value of the effective interactions; the extent to which the interactions (encoded in $A$) effectively happen in the dataset.
    \item $<A>_W$: the average of the inferred interaction matrix $A$ weighted by the effective interactions $W$. In this case, $W$ can be interpreted as our confidence in the corresponding entries of $A$.
    \item $\frac{<W^{intra}>}{<W^{extra}>}$: ratio of the intra-cluster effective interactions with the extra-cluster effective interactions; how much clusters self-interact versus how much they interact with other ones.
\end{itemize}

\begin{table}[h]
    \centering
    \caption{\textbf{Interaction strength} --- Overall, interaction between clusters is weak. The standard deviation on the last digits are given between parenthesis. The large standard deviations suggest that there is a large variety of interacting behaviours. Interactions tend to happen within a cluster (self-interactions).
    }
    \label{tab-interStrength}
    \begin{tabularx}{\textwidth}{|c|c|c|X|X|X|X|}
        \cline{1-7}
        $\vec{\kappa} (t)$ & $\theta_0$ & $r$ & $<A>$ ($10^{-3}$) & $<W>$ ($10^{-5}$) & $<A>_W$ ($10^{-3}$) & $\frac{<W^{intra}>}{<W^{extra}>}$ \\
        \cline{1-7}
        {\multirow{6}{*}{\rotatebox[origin=c]{90}{\textbf{Minute}}}} & {\multirow{3}{*}{\rotatebox[origin=c]{90}{\textbf{0.01}}}} & 0.5 & 49(21) & 342(889) & 66(17) & 1.8(62) \\
        & & 1.0 & 48(20) & 478(1124) & 60(17) & 1.4(43) \\
        & & 1.5 & 48(20) & 746(1901) & 60(17) & 1.0(33) \\
        \cdashline{2-7}
        & {\multirow{3}{*}{\rotatebox[origin=c]{90}{\textbf{0.001}}}} & 0.5 & 50(22) & 316(882) & 66(17) & 3.1(138) \\
        & & 1.0 & 50(21) & 279(752) & 67(16) & 2.6(105) \\
        & & 1.5 & 50(22) & 268(665) & 67(16) & 2.3(84) \\
        \cline{1-7}
        
        % ========================
        
        {\multirow{6}{*}{\rotatebox[origin=c]{90}{\textbf{Hour}}}} & {\multirow{3}{*}{\rotatebox[origin=c]{90}{\textbf{0.01}}}} & 0.5 & 49(18) & 389(843) & 56(17) & 0.5(13) \\
        & & 1.0 & 49(18) & 478(1187) & 56(17) & 0.6(15) \\
        & & 1.5 & 48(17) & 471(789) & 52(15) & 0.7(13) \\
        \cdashline{2-7}
        & {\multirow{3}{*}{\rotatebox[origin=c]{90}{\textbf{0.001}}}} & 0.5 & 50(21) & 110(398) & 61(17) & 1.7(67) \\
        & & 1.0 & 50(18) & 133(506) & 57(17) & 1.4(60) \\
        & & 1.5 & 49(17) & 183(554) & 55(17) & 1.1(37) \\
        \cline{1-7}
        
        % ========================
        
        {\multirow{6}{*}{\rotatebox[origin=c]{90}{\textbf{Day}}}} & {\multirow{3}{*}{\rotatebox[origin=c]{90}{\textbf{0.01}}}} & 0.5 & 49(18) & 41(97) & 55(17) & 1.2(34) \\
        & & 1.0 & 49(19) & 63(131) & 54(17) & 1.2(31) \\
        & & 1.5 & 49(19) & 91(187) & 53(18) & 1.2(31) \\
        \cdashline{2-7}
        & {\multirow{3}{*}{\rotatebox[origin=c]{90}{\textbf{0.001}}}} & 0.5 & 50(20) & 18(90) & 60(19) & 1.1(59) \\
        & & 1.0 & 50(19) & 23(101) & 58(17) & 1.0(50) \\
        & & 1.5 & 50(19) & 37(111) & 56(18) & 1.0(36) \\
        \cline{1-7}
         
    \end{tabularx}

\end{table}

\begin{figure}
    \centering
    \includegraphics[width=\textwidth]{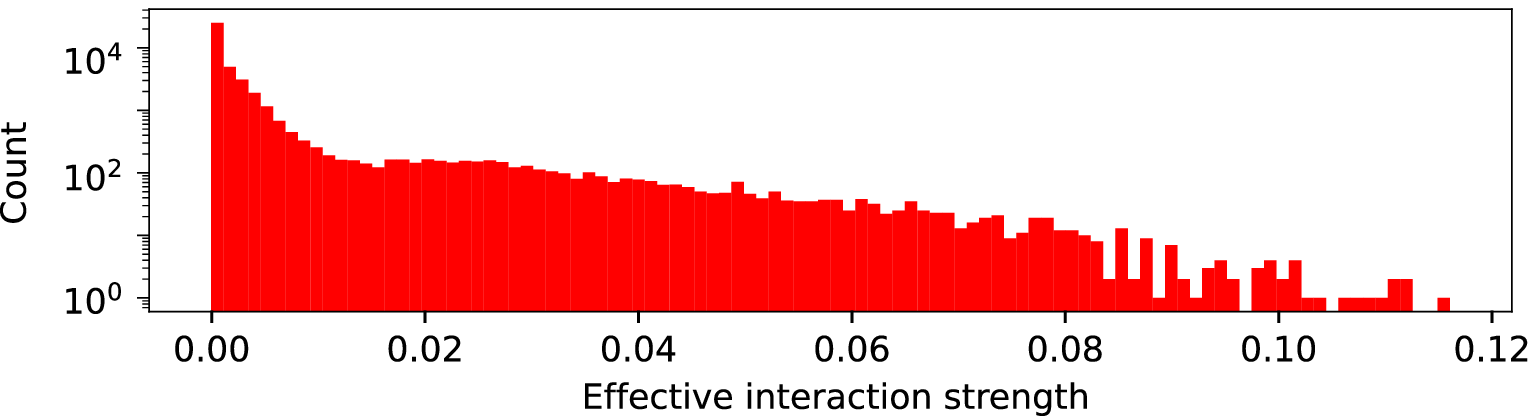}
    \caption{\textbf{Distribution of interaction strength} -- A large effective interaction strength means that the  publication of a new document can be significantly explained by the presence of previously published news. We see that most interactions are weak.}
    \label{fig-MPDHP-figinterstrength}
\end{figure}

%When computing the means, we discard the entries of $A$ and $W$ equal to 0. This is because not all clusters exist simultaneously, and thus should not be considered. An interaction strictly equal to 0 means that clusters simply did not exist at the same time.

The main conclusion of the results Table~\ref{tab-interStrength} is that most interactions are weak. The average value of $A$ tells us that the average value of the inferred parameters is around $0.05$, which is few given entries of $A$ are bounded between 0 and 1. The metric ${<W>}$ tells us that on all events, the interaction between clusters rose the probability of publication by 0.1\%-1\% on average. We can also note that the values of ${<W>}$ are of the same order of magnitude as $\lambda_0$ (0.01 for the ``Minute'' kernel, 0.001 for ``Hour'', and 0.0001 for ``Day''). We can interpret this as the probability for a new document belonging to a cluster or being from a new cluster is roughly the same from a temporal perspective. The metric $<A>_W$ tells us that when weighting the average of $A$ with the effective interaction, the values of $A$ are slightly higher than $0.05$; we can now be confident in this value, given it has been inferred on a statistically significant number of observations. Still, only some interactions seem to be significant, which correlates with \cite{Myers2012CoC,Poux2021IMMSBM}. Finally, the last metric $\frac{<W^{intra}>}{<W^{extra}>}$ finds that most effective interactions take place more within the same cluster; clusters tend to self-replicate. Further studies of this phenomena would involve an extension of the MPDHP that considers Nested Dirichlet Processes instead of Powered Dirichlet Processes. Clusters would then be broken down into smaller ones, whose interactions could be analyzed.
%Only for the ``Hour'' kernel and $\theta_0=0.01$ this value is lesser than one. It is because in this case, MPDHP infers two large clusters that exist for the entire year (see the 2 first rows of Fig.~\ref{fig-timeline-MPDHPNews}) and side topic-specific clusters. These clusters strongly influence each other, and all the topic-specific clusters can interact with them. In fact, the same effect explains the decrease of $\frac{<W^{intra}>}{<W^{extra}>}$ as $r$ grows: fewer clusters are inferred, and the probability of having large clusters that last for the whole period increases.

Another major observation from Table~\ref{tab-interStrength} is that standard deviations of effective interactions are large: it hints that some of interactions may play a more significant role in the dataset. In Fig.~\ref{fig-MPDHP-figinterstrength}, we plot the distribution of effective interactions for one specific run (``Hour'' kernel, $\theta_0=0.01$, $r=1$; we recover the same trend in all other experiments). The results of this figure are similar to the ones of previous studies \cite{Myers2012CoC,Poux2021IMMSBM}.

\subsubsection{Interactions range}
Finally, in Table~\ref{tab-interRange}, we investigate the range of effective interactions. We compute the effective interaction for each entry of the temporal kernel individually and average it over all existing clusters. 
Importantly, the effective interaction of $\kappa_1$ is smaller than others. This is induced by our kernel choice: because $\kappa_1$ is centered around $t=0$, half of the associated Gaussian function accounts negative time differences, which never happens by design. Therefore, only other kernels entries can contribute on both sides of their mean.

\begin{table}[t]    
    \centering
    \caption{\textbf{Interaction range} --- All the values for effective interaction are given in ten-thousandth ($10^{-5}$). Influence tends to decrease over time for all kernels.}
    \label{tab-interRange}
    \begin{tabularx}{\textwidth}{|c|c|c|X|X|X|X|X|X|X|X|X|X|}
        \cline{1-12}
        $\vec{\kappa} (t)$ & $\theta_0$ & $r$ & $\kappa_1$ & $\kappa_2$ & $\kappa_3$ & $\kappa_4$ & $\kappa_5$ & $\kappa_6$ & $\kappa_7$ & $\kappa_8$ & $\kappa_9$ \\
        \cline{1-12}
        & & & 0m & 10m & 20m & 30m & 40m & 50m & 60m & 70m & 80m \\
        \cline{1-12}
        {\multirow{6}{*}{\rotatebox[origin=c]{90}{\textbf{Minute (m)}}}} & {\multirow{3}{*}{\rotatebox[origin=c]{90}{\textbf{0.01}}}} & 0.5 & 133 & 421 & 407 & 451 & 428 & 403 & 395 & 345 & 121 \\
        & & 1 & 198 & 591 & 580 & 607 & 532 & 575 & 521 & 507 & 224 \\
        & & 1.5 & 308 & 937 & 893 & 914 & 955 & 840 & 808 & 810 & 304 \\
        \cdashline{2-12}
        & {\multirow{3}{*}{\rotatebox[origin=c]{90}{\textbf{0.001}}}} & 0.5 & 218 & 509 & 457 & 424 & 371 & 340 & 313 & 178 & 52 \\
        & & 1 & 142 & 435 & 396 & 388 & 343 & 327 & 272 & 187 & 45 \\
        & & 1.5 & 104 & 388 & 366 & 353 & 326 & 333 & 290 & 215 & 61 \\
        \cline{1-12}
        
        % ==================
        
        & & & 0h & 2h & 4h & 6h & 8h & - & - & - & - \\
        \cline{1-12}
        {\multirow{6}{*}{\rotatebox[origin=c]{90}{\textbf{Hour (h)}}}} & {\multirow{3}{*}{\rotatebox[origin=c]{90}{\textbf{0.01}}}} & 0.5 & 247 & 430 & 502 & 456 & 324 & & & & \\
        & & 1 & 329 & 538 & 549 & 542 & 451 & & & & \\
        & & 1.5 & 229 & 615 & 532 & 526 & 411 & & & & \\
        \cdashline{2-8}
        & {\multirow{3}{*}{\rotatebox[origin=c]{90}{\textbf{0.001}}}} & 0.5 & 62 & 149 & 119 & 137 & 92 & & & & \\
        & & 1 & 77 & 164 & 172 & 149 & 111 & & & & \\
        & & 1.5 & 104 & 244 & 223 & 197 & 156 & & & & \\
        \cline{1-12}
        
        % ==================
        
        & & & 0d & 1d & 2d & 3d & 4d & 5d & 6d & - & - \\
        \cline{1-12}
        {\multirow{6}{*}{\rotatebox[origin=c]{90}{\textbf{Day (d)}}}} & {\multirow{3}{*}{\rotatebox[origin=c]{90}{\textbf{0.01}}}} & 0.5 & 22 & 45 & 47 & 46 & 47 & 47 & 37 & & \\
        & & 1 & 35 & 71 & 72 & 68 & 68 & 70 & 61 & & \\
        & & 1.5 & 51 & 100 & 101 & 105 & 98 & 105 & 82 & & \\
        \cdashline{2-10}
        & {\multirow{3}{*}{\rotatebox[origin=c]{90}{\textbf{0.001}}}} & 0.5 & 9 & 20 & 21 & 21 & 21 & 21 & 17 & & \\
        & & 1 & 12 & 26 & 24 & 27 & 28 & 26 & 22 & & \\
        & & 1.5 & 11 & 41 & 42 & 41 & 41 & 41 & 35 & & \\
        \cline{1-12}

    \end{tabularx}
\end{table}

We see in Table~\ref{tab-interStrength} that influence tends to decrease over time for all the kernels considered. Overall, the interaction between documents still plays a marginal role. We did not plot the standard deviation for visualization purposes, but they are similar as in Table~\ref{tab-interStrength}; \textit{most} interactions do not play a significant role in the publication of subsequent documents over time. Overall, the increase in probability for a new document to belong to a cluster due to interactions is within 0.1\%-1\%.

\section{Conclusion}
In this work, we conducted extensive experiments on a real-world large-scale dataset from Reddit. We conducted 24 different experiments, each accounting for a given combination of parameters, that determine the timescale considered ($\kappa(\Delta t)$), the sparsity of the language modeling ($\theta_0$), and the extent on which we rely on text or time during the inference ($r$).

Our experiments hint that interactions do not play a significant role in this dataset. We proposed several ways to assess the role of interactions in the dataset. In particular, we introduced the notion of \textit{effective interaction} as a way to evaluate how confident we are in MPDHP's output. On this basis, we analysed the importance of interactions in general, as well as from a temporal perspective. We recovered the conclusions of prior works: interactions are sparse and decay over time. By looking at the global effective interaction average, we conclude that interactions play a minor role this dataset. Overall, they only increase the instantaneous probability for a new observation to appear by 1\%. Even the most extreme interactions seem to only increase this probability by 12\% top. 

However, despite intending our study as exhaustive, there is room for improvement in interaction modelling using MPDHP. In particular, there are two biases that we could not explore in this work. Firstly, the parameter $\lambda_0$ has been set according to a heuristic (so that a new cluster is opened with fifty percent chances when we are 95\% sure that it does not match the existing one). Its direct inference would robustify the approach. Nevertheless, this inclusion sounds challenging: $\lambda_0$ does not account for individual events realizations, but for Hawkes processes starts, whose inference is not a trivial extension. Another improvement would be to allow clusters to passively replicate --without the need for an interaction. We expect that this would boil down to adding a time-independent kernel entry to $\vec{\kappa}$. However, other questions may arise from such modification: when to consider a cluster as extinct given a non-fading kernel? How should this kernel relate to the temporal concentration parameter $\lambda_0$?
We believe such improvements would make MPDHP more robust and interpretable, and find applications beyond interaction modelling.

\bibliographystyle{splncs03}
\bibliography{1_Bibliography}

\end{document}